\documentclass[english,10pt,final,twocolumn]{elsarticle}
\usepackage[T1]{fontenc}
\usepackage[latin9]{inputenc}
\usepackage{pdfpages}
\usepackage{amstext}
\usepackage{graphicx}
\usepackage{gensymb}
\usepackage{amssymb}
\usepackage{placeins}
\usepackage{subfigure}
\usepackage[title]{appendix}
\makeatletter
\usepackage{float}

\usepackage{blindtext}
\usepackage{titlesec}

\titleformat{\section}
  {\bfseries\large}{\thesection}{1em}{}
\titleformat{\subsection}
  {\bfseries}{\thesubsection}{1em}{}
  \titleformat{\appendix}
  {\bfseries\large}{\thesection}{1em}{}

\newcommand{\lyxmathsym}[1]{\ifmmode\begingroup\def\b@ld{bold}
  \text{\ifx\math@version\b@ld\bfseries\fi#1}\endgroup\else#1\fi}
\usepackage{braket}
\usepackage{mathrsfs}
\usepackage{slashed}
\usepackage{bm}
\usepackage{datetime}
\usepackage{siunitx}

\usepackage{amsmath}

\DeclareSIUnit[number-unit-product = {}]\clight{c}
\DeclareSIUnit\eVperc{\eV\per\clight}
\DeclareSIUnit\GeVpercs{\giga\eV\squared\per\clight\squared}
\DeclareSIUnit\MeVpercs{\mega\eV\per\clight\squared}
\journal{Physics Letters B}
\biboptions{numbers,sort&compress}
\usepackage[left=1.6cm,right=1.04cm,top=1.65cm,bottom=1.65cm,columnsep=25pt]{geometry}
\usepackage{hyperref}  
\hypersetup{breaklinks = true, colorlinks = true, allcolors = magenta}
\usepackage{setspace}
\usepackage{graphicx}
\usepackage{wrapfig}

\newcommand\Tstrut{\rule{0pt}{2.6ex}}       
\newcommand\Bstrut{\rule[-1.2ex]{0pt}{0pt}} 

\usepackage{xcolor, soul}
\sethlcolor{yellow}

\makeatother

\usepackage{babel}
\begin{document}

\begin{frontmatter}{}

\title{Measurements of the electron-helicity asymmetry in the quasi-elastic ${\rm A}(\vec e,e' p)$ process}

\author[TAU]{T.~Kolar\corref{cor2}}
\ead{tkolar@mail.tau.ac.il}
\author[TAU]{S.J.~Paul\fnref{ucr}}
\author[Mainz]{P.~Achenbach}
\author[Mainz]{H.~Arenh\"ovel}
\author[TAU]{A.~Ashkenazi}
\author[JSI]{J.~Beri\v{c}i\v{c}}
\author[Mainz]{R.~B\"ohm}
\author[zagreb]{D.~Bosnar}
\author[JSI]{T.~Brecelj}
\author[Rutgers]{E.~Cline}
\author[TAU]{E.O.~Cohen\fnref{nrcn}}
\author[Mainz]{M.O.~Distler}
\author[Mainz]{A.~Esser}
\author[zagreb]{I.~Fri\v{s}\v{c}i\'{c}\fnref{mit}}
\author[Rutgers]{R.~Gilman}
\author[Pavia]{C.~Giusti}
\author[Konstanz]{M.~Heilig}
\author[Mainz]{M.~Hoek}
\author[TAU]{D.~Izraeli}
\author[Mainz]{S.~Kegel}
\author[Mainz]{P.~Klag}
\author[nrc,TAU]{I.~Korover}
\author[TAU]{J.~Lichtenstadt}
\author[TAU,soreq]{I.~Mardor}
\author[Mainz]{H.~Merkel}
\author[Mainz]{D.G.~Middleton}
\author[UL,JSI,Mainz]{M.~Mihovilovi\v{c} }
\author[Mainz]{J.~M\"uller}
\author[Mainz]{U.~M\"uller}
\author[TAU]{M.~Olivenboim}
\author[TAU]{E.~Piasetzky}
\author[Mainz]{J.~Pochodzalla}
\author[huji]{G.~Ron}
\author[Mainz]{B.S.~Schlimme}
\author[Mainz]{M.~Schoth}
\author[Mainz]{F.~Schulz}
\author[Mainz]{C.~Sfienti}
\author[UL,JSI]{S.~\v{S}irca}
\author[Mainz]{R.~Spreckels}
\author[JSI]{S.~\v{S}tajner }
\author[Mainz]{Y.~St\"ottinger}
\author[USK]{S.~Strauch}
\author[Mainz]{M.~Thiel}
\author[Mainz]{A.~Tyukin}
\author[Mainz]{A.~Weber}
\author[TAU]{I.~Yaron}
\author{\\\textbf{(A1 Collaboration)}}

\cortext[cor2]{Corresponding author}
\fntext[ucr]{Equal contributing author,\\\indent\hspace{1.3mm} Present address: UC-Riverside, Riverside, CA 92521, USA.}
\fntext[mit]{Present address: MIT-LNS, Cambridge, MA 02139 USA.}
\fntext[nrcn]{Present address: NRCN, Beer-Sheva 84190, Israel.}

\address[TAU]{School of Physics and Astronomy, Tel Aviv University, Tel Aviv 69978,
Israel.}
\address[Mainz]{Institut f\"ur Kernphysik, Johannes Gutenberg-Universit\"at, 55099
Mainz, Germany.}
\address[JSI]{Jo\v{z}ef Stefan Institute, 1000 Ljubljana, Slovenia.}
\address[zagreb]{Department of Physics, University of Zagreb, HR-10002 Zagreb, Croatia.}
\address[Rutgers]{Rutgers, The State University of New Jersey, Piscataway, NJ 08855,
USA.}

\address[Pavia]{Dipartimento di Fisica, Universit\`a degli Studi di Pavia and INFN, Sezione di Pavia, via A.~Bassi 6, I-27100 Pavia, Italy.}
\address[Konstanz]{Universit\"at Konstanz, Fachbereich Physik, Universit\"atsstra{\ss}e 10, 78464 Konstanz, Germany.}
\address[nrc]{Department of Physics, NRCN, P.O. Box 9001, Beer-Sheva 84190, Israel.}
\address[soreq]{Soreq NRC, Yavne 81800, Israel.}
\address[UL]{Faculty of Mathematics and Physics, University of Ljubljana, 1000
Ljubljana, Slovenia.}
\address[huji]{Racah Institute of Physics, Hebrew University of Jerusalem, Jerusalem
91904, Israel.}
\address[USK]{University of South Carolina, Columbia, South Carolina 29208, USA.}

\begin{abstract}
We present measurements of the electron helicity asymmetry in quasi-elastic proton knockout from $^{2}$H and $^{12}$C nuclei by polarized electrons. This asymmetry depends on the fifth structure function, is antisymmetric with respect to the scattering plane, and vanishes in the absence of final-state interactions, and thus it provides a sensitive tool for their study. Our kinematics cover the full range in off-coplanarity angle $\phi_{pq}$, with a polar angle $\theta_{pq}$ coverage up to about 8$\degree$. The missing energy resolution enabled us to determine the asymmetries for knock-out resulting in different states of the residual $^{11}$B system.  We find that the helicity asymmetry for $p$-shell knockout from $^{12}$C depends on the final state of the residual system and is relatively large (up to $\approx 0.16$), especially at low missing momentum. It is considerably smaller (up to $\approx 0.01$) for $s$-shell knockout from both $^{12}$C and $^2$H.  The data for $^2$H are in very good agreement with theoretical calculations, while the predictions for $^{12}$C exhibit  differences with respect to the data.

\end{abstract}
\date{\today}

\end{frontmatter}{}

\section{Introduction}
The large amount of data on the $(e,e'p)$ knockout reaction collected over the last decades in different laboratories, provided accurate information on the properties of proton-hole states \cite{Frullani:1984nn,vandersteen:88a,vandersteen:88b,Boffi:1993gs,Boffi:1996ikg}. The separation energy and the momentum distribution of the removed proton, which allows one to determine the associated quantum numbers, were obtained. From the comparison between the experimental and theoretical cross sections, it was possible to extract the spectroscopic factors, which are a measure of the occupation of the various shells \cite{vandersteen:88a,vandersteen:88b}.

It is well known \cite{Boffi:1996ikg,Giusti:1989ww,BOFFI1985697,Boffi:1982mr,vdsluyis:94} that the unpolarized cross section does not provide any detailed information on the various components of the scattering matrix. Thus for a more complete analysis of the process under study one needs additional information as provided by measurements of polarization observables which contain the scattering components of the response of the nucleus to the electromagnetic probe in different combinations. In particular, small components which might contain important information on dynamical details of the process and are buried in the unpolarized cross section, may become accessable in some polarization observables. Their determination imposes severe constraints on theoretical models and is therefore of great interest \cite {vdsluyis:94}.

When the incoming-electron beam is longitudinally polarized  with beam polarization $P_e$ and a given helicity  $h$, the $(\vec e,e'p)$ cross section can be written, in the one-photon-exchange approximation, as the sum of a helicity-independent term  $\Sigma$ (the unpolarized cross section) and a helicity-dependent term $hP_e\Delta$ \cite{Boffi:1996ikg,Giusti:1989ww,BOFFI1985697}:
\begin{equation} 
\begin{aligned}
\sigma = &C(\rho_L f_L+\rho_T f_T+\rho_{LT} f_{LT}\cos \phi_{pq}\hspace{1.3cm}\\
&+\rho_{TT}f_{TT}\cos 2\phi_{pq}+h P_e\rho'_{LT}f'_{LT}\sin\phi_{pq})\\  
= &\Sigma + hP_e\Delta,
\end{aligned}
\label{eq:xs}
\end{equation}
where $C$ and the $\rho$'s depend only on the electron kinematics, and $\phi_{pq}$ is the off-coplanarity angle (see Fig.~\ref{fig:kinematics_diagram}). Here, the unpolarized cross section is written as a combination of four structure functions ($f_L,f_T,f_{LT},f_{TT}$). The structure functions represent the response of the nucleus to the various components of the electromagnetic charge and transverse current components and can be expressed as bilinear combinations of scattering amplitudes \cite{Boffi:1996ikg,Giusti:1989ww,Boffi:1982mr,vdsluyis:94}. The helicity-dependent term is governed by the fifth structure function, $f'_{LT}$, \cite{Boffi:1996ikg,Giusti:1989ww,BOFFI1985697}, and is asymmetric with respect to the scattering plane. To observe this asymmetry one needs to detect the emitted proton in out-of-plane kinematics ($\phi_{pq} \neq 0$) while flipping the electron helicity. The fifth structure function is the imaginary part of the longitudinal-transverse interference component of the hadron tensor. In the plane-wave impulse approximation (PWIA), where the final-state interaction (FSI) between the outgoing proton and the residual nucleus is neglected, the hadron tensor is real and the fifth structure function vanishes identically \cite{Boffi:1996ikg,BOFFI1985697,Donnelly:1986aq}. Therefore the fifth structure function and the helicity asymmetry is determined by FSI and thus provides a critical test of any theoretical description.  

A series of measurements of the ${\rm A}(\vec e,e'\vec p)$ reaction was performed at the Mainz Microtron (MAMI) using $^2$H and $^{12}$C targets. The polarization transfer to the knock-out proton was measured to investigate the properties of deeply bound (highly off-mass-shell) protons in the nucleus  \cite{deep2012PLB,deepCompPLB,posPmissPLB,ceepLet,ceepComp,ceepTim}. In search of deviations of the form factors of a bound proton from the free proton ones these measurements have been compared to calculations which are sensitive to this ratio. The treatment of the reaction mechanism and the nuclear effects in these calculations must be tested against experiments which are less sensitive to the nucleon structure, such as the induced polarization in the knocked-out proton by the ${\rm A}(e,e'\vec p)$ reaction \cite{induced} or the electron-helicity asymmetry through the ${\rm A}(\vec e,e'p)$ reaction.

We report here the measurement of the analyzing power of the electron beam in these datasets,
which is also referred to as the electron-helicity asymmetry $A$, obtained from the
$(\vec e,e'p)$ reaction on $^2$H and $^{12}$C, where only the incoming electron beam is
polarized. This asymmetry is defined as the ratio of the helicity-dependent to the helicity-independent part of the cross section, $\Delta \over \Sigma$. It is therefore proportional to the fifth structure function $f'_{LT}$, which is governed by FSI as outlined above. Earlier measurements of $A$ were performed at the  MIT Bates Linear Accelerator Center \cite{PhysRevLett.72.3325,Dolfini:1995zz,Dolfini:1999sk} on   $^2$H and $^{12}$C. For $^{12}$C, only $p$-shell protons were measured, but without distinction between the final states of the $^{\rm 11}$B residual system. 

Our measurements cover the full 360$\degree$ range of the off-coplanarity angle $\phi_{pq}$  showing the sin($\phi_{\rm pq}$) behavior, and  have much better statistics and energy resolution
than the previous data. For the protons knocked out from the $^{12}\mathrm{C}$ $p$ shell, the data allowed us to extract separate asymmetries for transitions to a few bound states of the residual system---$^{\rm 11}$B ground state and two excited states---and show that the underlying structure functions are different. We obtained the asymmetries also for protons knocked out from the $^{12}$C $s$ shell, where the residual system exhibits a continuous spectrum. The smaller uncertainties enable a meaningful comparison to state-of-the-art calculations. The measurements cover a relatively large range in the ``missing momentum'', $p_{\rm miss}$, (which in the impulse approximation and the absence of FSI corresponds to the negative initial momentum of the knocked-out proton). The data allowed to study the dependence of $A$  on two kinematic variables: the angle $\theta_{pq}$ between the emitted proton and the momentum transfer (see Fig.~\ref{fig:kinematics_diagram}),  and the missing momentum $p_{\rm miss}$.

\section{Experimental setup and kinematics}
The experiments were performed at the  MAMI facility of the Johannes Gutenberg University at Mainz using the A1 beamline and spectrometers \cite{a1aparatus}.  For these measurements, a polarized electron beam of 600, 630, and 690 MeV was used.  The beam current was $\approx 10$ $\mu$A, with an $\approx 80-89\%$ polarization. The beam helicity was flipped at a rate of 1 Hz.  

The beam polarization was measured periodically with a M\o ller polarimeter \cite{Wagner,Bartsch}, and verified by a Mott polarimetry measurement \cite{Tioukine}.  These two methods of beam-polarization measurement produced mutually consistent results  as described in \cite{deepCompPLB,posPmissPLB,ceepComp,ceepTim}.

\begin{figure}[hb!]
\includegraphics[width=\columnwidth]{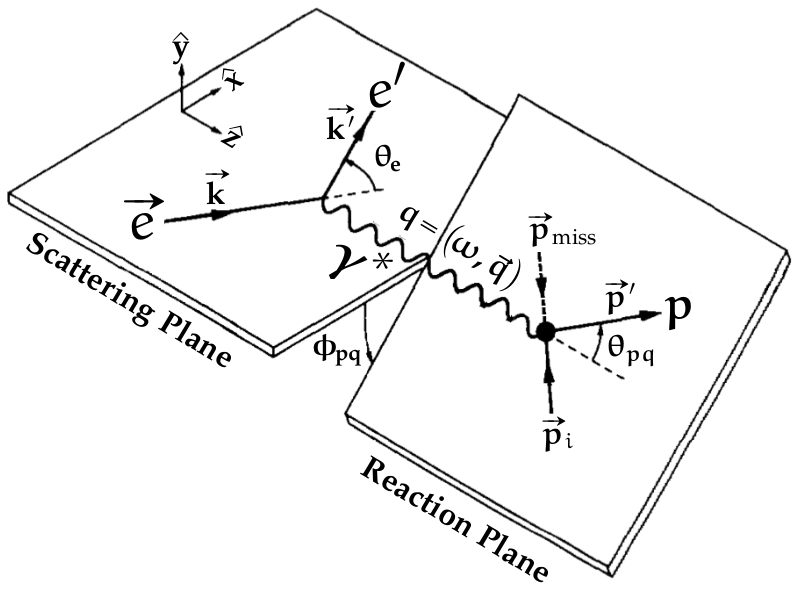}
\caption{
Kinematics of the $(\vec{e},e^{\prime},p)$ reaction with the definitions of the kinematic variables.
}
\label{fig:kinematics_diagram}
\end{figure} 
 
\begin{table*}[t!]
\caption{
The kinematic settings in the $^{2}{\rm H}(\vec{e},e' p\,)$ and $^{12}{\rm C}(\vec{e},e' p\,)$ measurements presented in this work. The angles and momenta represent the central values for the two spectrometers: $p_p$ and $\theta_p$ ($p_e$ and $\theta_e$) are the knocked out proton (scattered electron) momentum and scattering angles, respectively.   
}
\begin{center}
\begin{tabular}{llllll}
\hline
\multicolumn{2}{c}{}& \multicolumn{4}{l}{Kinematic setting}  \Tstrut\Bstrut\\ \cline{3-6}
 & & C & F & A & G \Tstrut\\
\hline
Target & & $^2\mathrm{H}$& $^2\mathrm{H}$& $^{12}\mathrm{C}$& $^{12}\mathrm{C}$ \Tstrut \\
$E_{\rm beam}$ & [MeV] &  630 & 690 & 600 & 600\\
$Q^2$ & [$({\rm GeV}\!/\!c)^2$] & 0.18 & 0.40 & 0.40 & 0.18\\
$p_{\rm miss}$ & [MeV$\!/\!c$]  & $-$82 to $-24$ & $-$90 to 71 & $-$110 to 96  & $-$250 to $-$120\\
$p_e$ & [MeV$\!/\!c$] & 509 &  474 & 384 & 368 \\
$\theta_e$ & [deg] & 43.4   & 67.1 & 82.4 & 52.9\\
$p_p$ & [MeV$\!/\!c$]  & 484   & 668 & 668 & 665\\
$\theta_{p}$ & [deg] & $-$53.3 &  $-$40.8 & $-$34.7 & -37.8\\
\hline
\end{tabular}
\end{center}
\label{tab:kinematics}
\end{table*}

The target used for the $^2$H measurements was a 50 mm long oblong cell filled with liquid deuterium \cite{deep2012PLB,deepCompPLB,posPmissPLB} while the $^{12}$C target consisted of three 0.8 mm-thick graphite foils \cite{ceepLet,ceepComp,ceepTim}.  We also performed calibration runs using a liquid-hydrogen target.  
 
Two high-resolution spectrometers with momentum acceptances of 20-25\% were used to detect the scattered electrons and knocked-out protons in coincidence.  Each of these spectrometers consists of a momentum-analysing magnet system followed by a set of vertical drift chambers for tracking, and a scintillator system for triggering and defining the time coincidence between the two spectrometers.  
 
The electron spectrometer was equipped with a \v Cerenkov detector in order to distinguish the scattered electrons from other types of particles. More details of the experiment can be found in \cite{deep2012PLB,deepCompPLB,posPmissPLB,ceepLet,ceepComp,ceepTim}.  
 
The kinematics of the measured reactions are shown in Fig.~\ref{fig:kinematics_diagram}.  The electron's initial and final  momenta are $\vec k$ and $\vec k\,'$ respectively, which define the scattering plane of the reaction.  The reaction plane is defined by the momentum transfer $\vec q = \vec k-\vec k\,'$ and the recoiling proton's momentum $\vec p\,'$.  We refer to the angle between the scattering plane and the reaction plane as the ``off-coplanarity'' angle of the reaction, denoted by $\phi_{pq}$.  
 
The missing momentum  $\vec p_{\rm miss}\equiv \vec q - \vec p\,'$ is the recoil momentum of the residual nuclear system. Neglecting FSI, $-\vec p_\mathrm{miss}$ is equal to the initial momentum of the emitted proton, $\vec p_i$. We conventionally define positive and negative signs for $p_{\rm miss}$ by the sign of $\vec p_{\rm miss}\cdot\vec q$.

In this work, we present measurements for two kinematic settings for each nucleus.  In both cases, there was one kinematic setting with $Q^2=0.18\,($GeV$/c)^2$ with negative missing momentum, and another  with $0.40\,($GeV$/c)^2$, with missing momentum centered at zero.  The details of these kinematic settings are given in Table \ref{tab:kinematics}.

\section{Analysis}
In addition to the cut on coincidence time between the two spectrometers, which reduces the amount of random coincidences, particle-identification and event-quality software cuts were applied to the data\footnote{The cuts applied here are similar to those used in polarization-transfer analysis, but without the cuts necessary for the measurement of the outgoing-proton polarization (e.g. proton-polarimeter and spin-precession cuts). This results in much higher event tally compared to those measurements \cite{deep2012PLB,deepCompPLB,posPmissPLB,ceepLet,ceepComp,ceepTim,induced}.}.  

The helicity asymmetry is evaluated as 
\begin{equation}
A = \frac{1}{P_e}\left( \frac{N_+ - N_-}{N_++N_-}\right)  = \frac{\Delta}{\Sigma} ,
\end{equation}
where $N_+$ ($N_-$) is the number of events which are measured  when the beam has positive (negative) helicity and $P_e$ is the measured beam polarization.  The raw measurements of $A$ must be  checked and corrected for  possible small discrepancies (on the order of 0.1\%) between the number of incident electrons with positive and negative helicity.  We correct for this bias by subtracting an offset $A_{0}$ from the measured value of $A$, where $A_{0}$ is determined from the best fit of the data to the function
\begin{equation}
    A_{\rm measured}=A\sin\phi_{pq}+A_{0}\,,
\end{equation}
neglecting a small $\cos\phi_{pq}$ and $\cos(2\phi_{pq})$ dependencies in $A$. The uncertainty of the fitted parameter $A_{0}$ was included in the systematic error. As a quality assurance check, we found that the same values of $A_{0}$ were obtained (within error) by selecting near-coplanar events only, that is, events within $\pm 1 \degree$  of $\phi_{pq}$ near 0$\degree$, $+180\degree$, and $-180\degree$ (where $A$ is expected to vanish), and evaluating $A$ for this subsample. 

Following \cite{deep2012PLB,deepCompPLB,posPmissPLB,induced}, we required the missing mass of the $^2$H$(\vec e,e' p\,)$ reaction to be consistent with the mass of a neutron. For the $^{12}$C sample, we distinguish between protons knocked out from the $s$ and $p$ shells, following \cite{vandersteen:88b,Dutta,ceepComp,ceepTim,induced}, by using cuts on the missing energy of the reaction, $E_{\rm miss}$, defined as  \cite{Dutta}:
\begin{equation}
	E_{\rm miss} \equiv \omega - T_p - T_{^{11}\rm B},
\end{equation}
where $\omega=k^0-k'^{\,0}$ is the energy transfer, $T_p$ is the measured kinetic energy of the outgoing proton, and $T_{^{11}\rm B}$ is the calculated kinetic energy of the recoiling residual system, assuming it is $^{11}$B in the ground state. For the $s$-shell sample, we used the cut $30< E_{\rm miss} < 60$ MeV,  while for the $p$-shell sample, we used  $15< E_{\rm miss} < 25$ MeV \cite{vandersteen:88b,Dutta,ceepLet,ceepComp,ceepTim,induced}.

The $p$-shell cut accepts events in which the residual $A-1$ system is left in one of several discrete states, including the ground-state of $^{11}$B as well as a few excited states. The $s$-shell selection cut is much wider, comprising a broad range within the continuum of unbound residual $A-1$ states.  

\begin{figure}[h]
\includegraphics[width=\columnwidth]{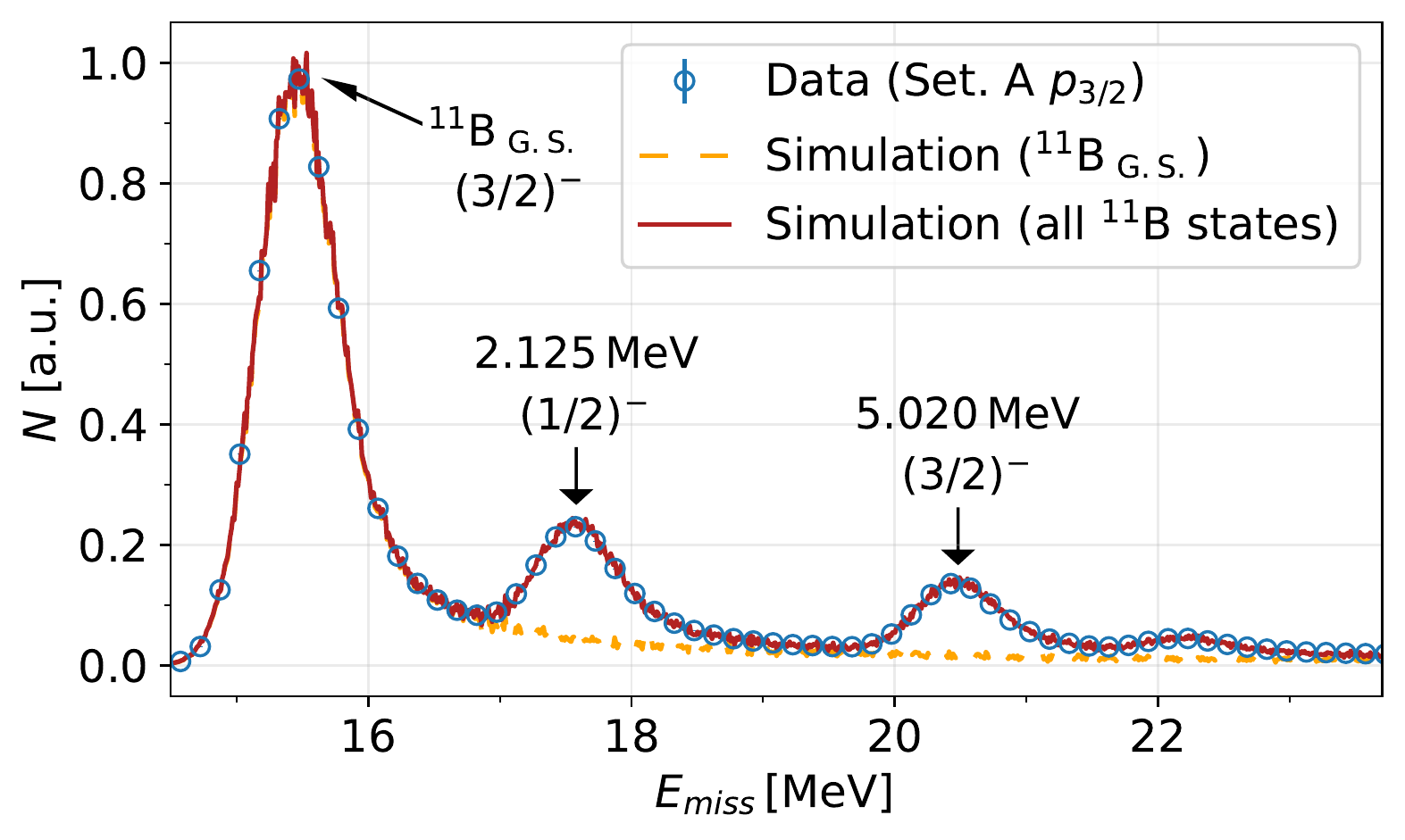}
\caption{
Missing-energy spectrum of the $p$-shell knockout from the $^{12}\mathrm{C}$ nucleus. The observed peaks correspond to different excited states of the $^{11}\mathrm{B}$ nucleus. Lines show a Monte-Carlo simulation obtained from the $^{12}\mathrm{C}$ structure function which includes an estimation of the energy losses. 
}
\label{fig:emiss_histogram}
\end{figure}

Previous measurements obtained the helicity asymmetry for $p$-shell knocked-out protons with no distinction of the final state of the residual system \cite{PhysRevLett.72.3325,Dolfini:1995zz,Dolfini:1999sk}. Since the structure function represents the nuclear response to the electromagnetic interaction, the helicity asymmetry may well depend on the final state of the residual system. The measured  $E_{\rm miss}$ spectrum for the low $p_{\rm miss}$ data (set A)  in our experiment, shown in Fig.~\ref{fig:emiss_histogram},  has sufficient resolution to separate events of different final states of the residual system, corresponding to the $^{11}$B ground state ($J^{\pi} = 3/2^{-}$), the first excitation at  2.125 MeV ($J^{\pi} = 1/2^{-}$) and the excitation at 5.020 MeV ($J^{\pi} = 3/2^{-}$). The unresolved excitation at 4.445 MeV is expected to be very weak \cite{vandersteen:88a,vandersteen:88b}. However, the raw asymmetry obtained from all events within the $E_{\rm miss}$ boundaries of an excited state is an average (over the number of events) of asymmetries of the events resulting in the excited state and contributions from events in the radiative tail of the lower lying states that extend under the chosen excitation peak.

We used a Monte-Carlo simulation using the $^{12}{\rm C}$ structure function and embedded target energy losses for the determination of the shape of the peak and its radiative tail. Fitting this lineshape to the measured peak enabled us to determine the number of the events under the peak corresponding to the first excited state that belong to the radiative tail of the elastic peak. The true asymmetry of the 'excitation-events', $A_{\rm Ex}$, is then determined by solving
\begin{equation}
 A_{\rm Tot} = \frac{ N_{\rm rad} A_{\rm GS} + N_{\rm Ex} A_{\rm Ex} }{N_{\rm Tot}} ,
\end{equation}
where $N_{\rm rad}$, $N_{\rm Ex}$, and $N_{\rm Tot}$ are the number of events in the $E_{\rm miss}$ region of the excitation due to the radiative tail, the events leading to the chosen excitation, and the total number of events, respectively. Similarly, the $A_{\rm GS}$, $A_{\rm Ex}$ and $A_{\rm Tot}$ are asymmetries of the ground-state events, the excitation events, and their measured average. The procedure was repeated iteratively  for the second excitation peak considering the tails of the ground state  and the first excitation peaks with their corresponding asymmetries.

There are several sources of systematic uncertainty in the helicity asymmetry, and they are summarized in Table \ref{tab:systematics}. The largest contribution is the uncertainty of the beam polarization ($\approx 2\%$) described in \cite{deepCompPLB,posPmissPLB,ceepComp,ceepTim}, which contributes a relative uncertainty of the same size to the helicity asymmetry.  Secondly, there is an uncertainty due to $A_{0}$ that is subtracted from the data.  This uncertainty is $\approx 0.0001$ for the carbon datasets, and $\approx 0.0003$ for the deuteron datasets.  

The systematic uncertainty introduced through the software cuts on different variables was determined by slightly tightening each cut individually and re-performing the analysis for each bin in $\phi_{pq}$. To get the total systematic error from the software cuts we took the root-mean-square of these contributions. For the $^2$H ($^{12}$C) datasets, this yields 0.0003 (0.0011), with the largest contribution coming from the missing-mass (missing-energy) cuts. With coincidence-time cut the effect of random coincidence events on the asymmetry is negligible (of the order $0.1\%\times A$). 

There are also systematic uncertainties attributed to the kinematic settings of the detector system that could result in bin-migration. To determine the effect of this uncertainty on our results, we re-performed the analysis with each of the following kinematic variables modified from the nominal values by $\pm$ the uncertainty thereof: $E_{\rm beam}$, $p_{e}$, $\theta_{e}$, $p_{p}$, $\theta_{p}$.  

\begin{table}[h]
    \centering
    \begin{tabular}{l|rr}
        Source & \multicolumn{2}{c}{$\Delta A$} \\
        \hline
        & $^2$H & $^{12}$C\\
        Beam polarization & $2\%\times A$ & $2\%\times A$ \\
        $A_{0}$ & 0.0003 & 0.0001 \\
        Software cuts & 0.0003 & 0.0011 \\
        Kinematics & 0.0003 & 0.0048 \\
        Total ($^2$H) & \multicolumn{2}{c}{$\sqrt{0.0005^2+(2\%A)^2}$}\\
        Total ($^{12}$C) & \multicolumn{2}{c}{$\sqrt{0.0049^2+(2\%A)^2}$}\\
    \end{tabular}
    \caption{Sources of systematic uncertainties.}
    \label{tab:systematics}
\end{table}

\section{Electron-helicity asymmetries}

\begin{figure}[t!]
	\center{\large $^2{\rm H}(\vec e,e'p)$}
	\includegraphics[width=\columnwidth]{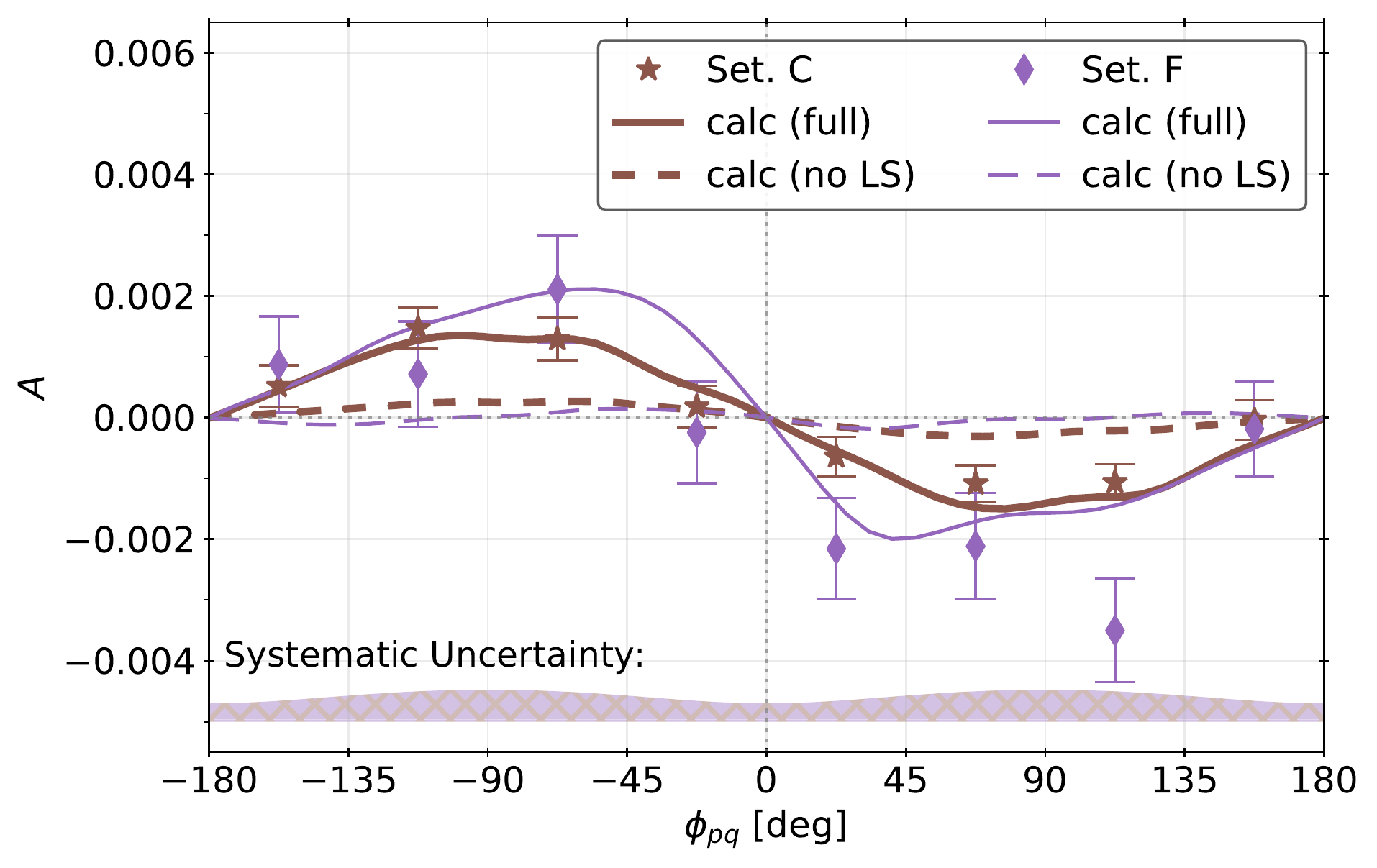}
	\caption{
		Helicity asymmetry, $A$, as a function of $\phi_{pq}$ for $^2$H,
		for kinematic settings C and F, with the statistical (bars) and the systematic uncertainties (shaded). Theoretical calculations with (without) the $L\cdot S$ interaction are shown as solid (dashed) curves.
	}
\label{fig:phid}
\vspace{0cm}
	\center{\large $^{12}{\rm C}(\vec e,e'p)$}
	\includegraphics[width=\columnwidth]{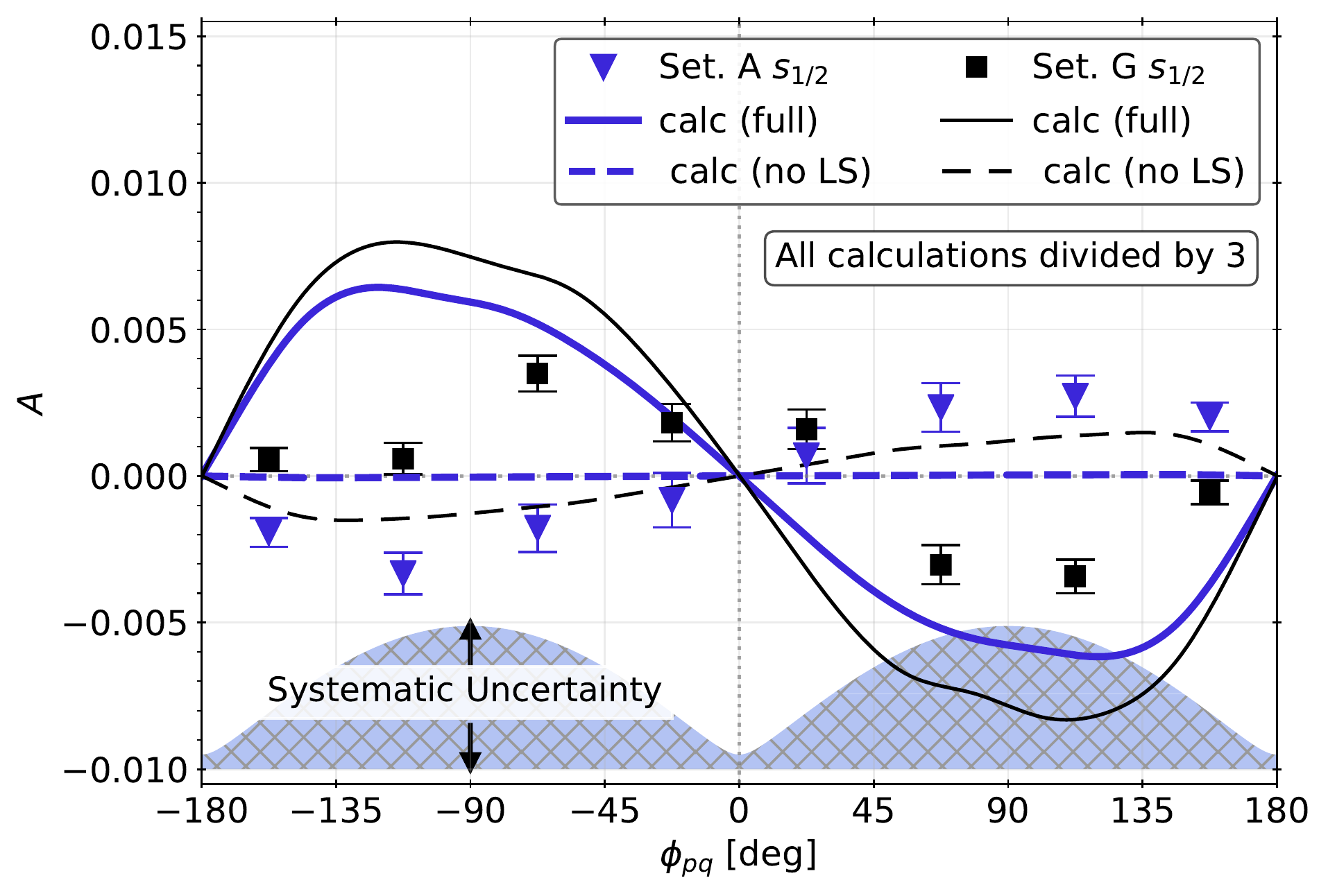}
	\caption{
		Helicity asymmetry, $A$, as a function of $\phi_{pq}$ for protons knocked out from the $s$ shell in $^{12}$C for low (Set A) and high (Set G) $p_{\rm miss}$ kinematic settings. Theoretical calculations with (without) the $L\cdot S$ interaction are multiplied by a factor $1/3$ and shown as solid (dashed) curves.
	}
	\label{fig:phis}
\end{figure}

\begin{figure}[ht!]
	\center{\large $^{12}{\rm C}(\vec e,e'p)$}
	\includegraphics[width=\columnwidth]{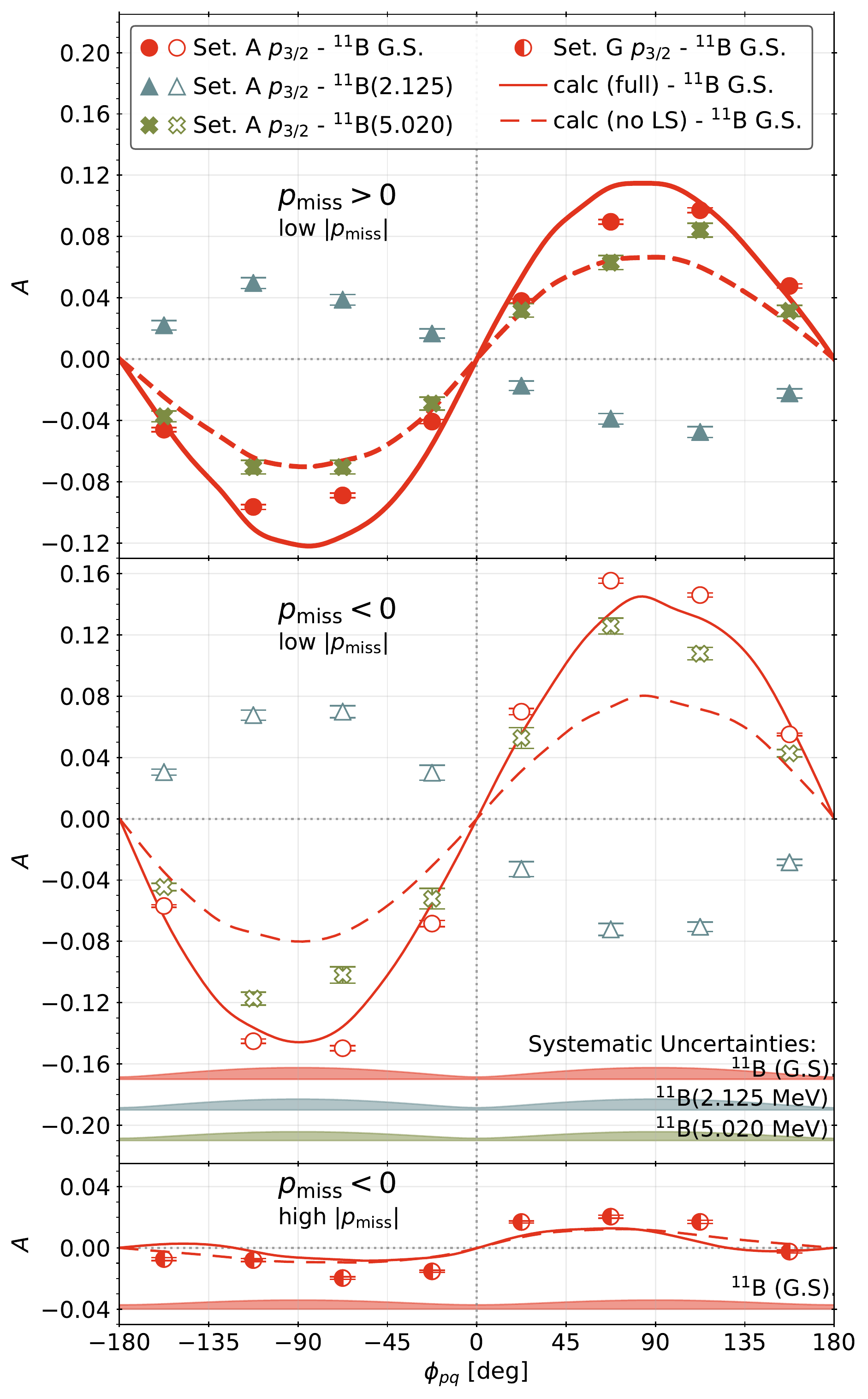}
	\caption{
		Helicity asymmetry, $A$, plotted as a function of $\phi_{pq}$ for protons knocked
		out from the $p$ shell in $^{12}$C. For the low-$p_{\rm miss}$ setting, asymmetries are shown for different states of the residual system: $^{11}$B ground state, 2.125 Mev (J$^{\pi}$=1/2$^{-}$), and 5.020 $(J^{\pi}$=3/2$^{-}$) excitations. The asymmetries are shown separately for positive (top) and negative $p_{\rm miss}$ (middle). For the high-$p_{\rm miss}$ setting the asymmetry with $^{11}$B ground state is shown (bottom). Theoretical calculations with (without) the $L\cdot S$ part of the interaction are shown as solid (dashed) curves.
	}
	\label{fig:phip}
	\vspace{0cm}
\end{figure}

The measured helicity asymmetries as a function of $\phi_{pq}$ are shown for $^2$H  in Fig.~\ref{fig:phid}. Asymmetries for knocked-out protons  from the $s$ shell in  $^{12}$C,  are shown in Fig. \ref{fig:phis} and from the $p$ shell  in Fig. \ref{fig:phip}. 

They show an overall $\sin(\phi_{pq})$ dependence, as expected from Eq.~\ref{eq:xs}, albeit with some distortion in shape which may be due to dependence on $\phi_{pq}$ in the helicity independent terms in $\Sigma$ that constitute the measured asymmetry. The data are compared to theoretical predictions which are discussed in Sec.~\ref{sec:calc}.

\subsection {Protons from $^{2}\mathrm{H}$ and the $s$ shell of $^{12}\mathrm{C}$.}
The measured asymmetries for the protons in  $^2$H shown in Fig.~\ref{fig:phid} (primarily in $s$ state) are relatively small (on the order of 0.001) and the dependence on $\phi_{pq}$ follows a sine with a negative amplitude.

In $^{12}$C, the asymmetries for the protons  ejected from the $s$ shell shown in Fig.~\ref{fig:phis} are also very small, similar in size to those in $^{2}$H. Previous measurements on $^{12}$C lacked sufficient statistics in this missing-energy range \cite {Dolfini:1999sk}. The low-$p_{\rm miss}$ region indicates a sine dependence with a positive amplitude, while in the high-$p_{\rm miss}$ region the amplitude is negative. 

We note that after the $s$-shell proton knockout the state of residual system is part of the continuum. Since the fifth structure function which underlies the asymmetry depends on the quantum numbers of the residual system (as demonstrated in the measurement of proton ejection from the  the $p$ shell  with different residual nuclear systems), asymmetries with opposite signs may contribute resulting in a reduced average asymmetry. Furthermore, variation in the cross section of the different excitations in the residual system will change the contribution to the measured (average)  asymmetry, which may explain the different phase observed at low and high $p_{\rm miss}$.
	
\begin{figure}[b!]
    \center{\large $^2{\rm H}(\vec e,e'p)$}
	\includegraphics[width=\columnwidth]{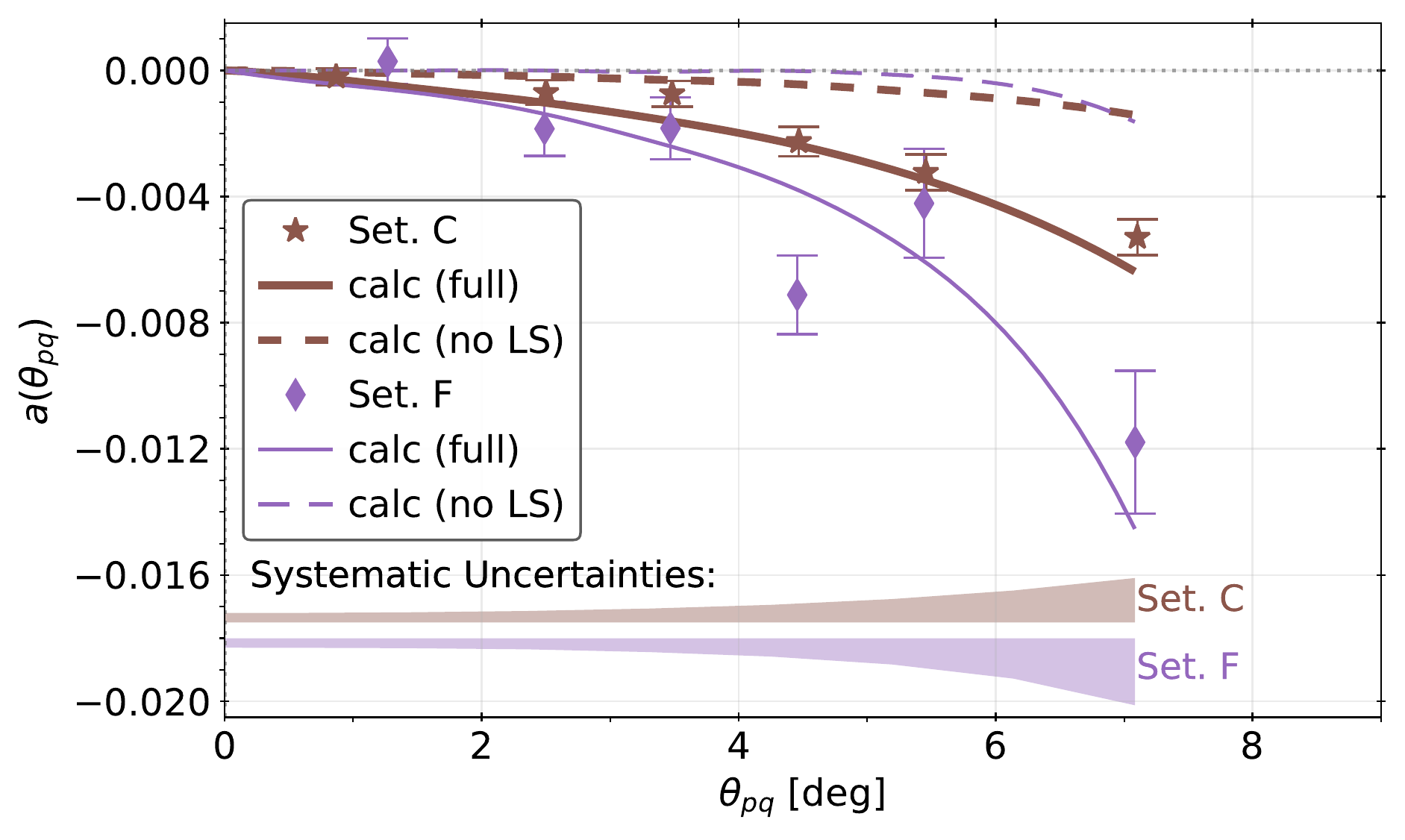}
	\center {\large $^{12}{\rm C}(\vec e,e'p)$}
	\includegraphics[width=\columnwidth]{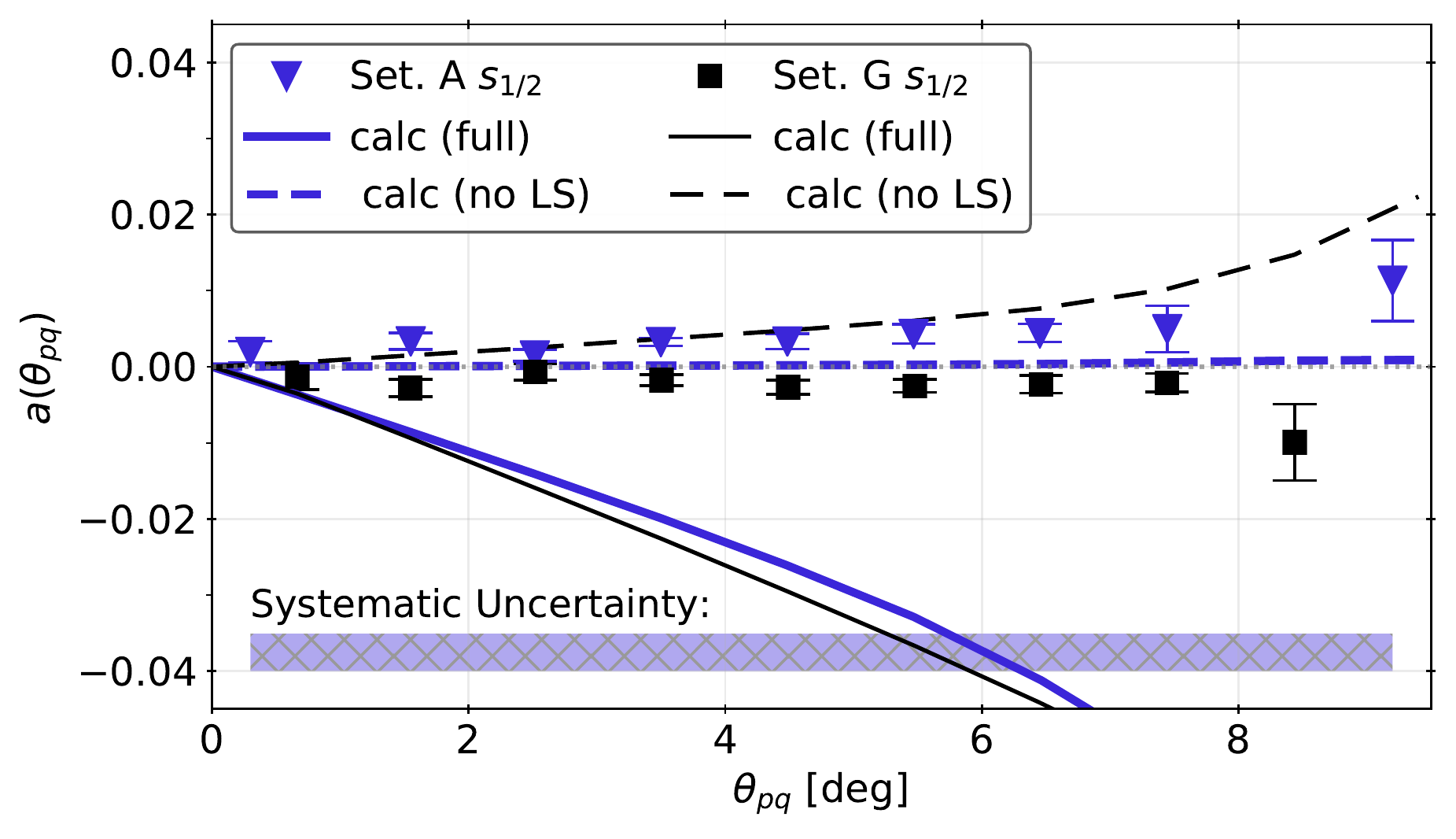}
	\caption{
		The fitted helicity-asymmetry amplitude in individual $\theta_{pq}$ bins, $a(\theta_{pq})$, for $^2$H (top panel) and $s$-shell knockout from $^{12}$C (bottom panel). The curves represent theoretical calculations with (solid) and without (dashed) $L\cdot S$ coupling.   
	}
	\label{fig:theta}
\end{figure}

\subsection{Protons from the $p$ shell in $^{12}\mathrm{C}$}
The asymmetries for the $p$-shell protons in $^{12}$C are shown in Fig.~\ref{fig:phip}, for the low (top two panels) and high (bottom panel)  $p_{\rm miss}$ settings. For the low-$p_{\rm miss}$ kinematics, the $p$-shell asymmetries are separated for transitions to the $^{11}{\rm B}$ ground state and the excitation to the  2.125 MeV (J$^{\pi}$ = 1/2$^{-}$) or 5.020 (J$^{\pi}$ = 3/2$^{-}$) $^{11}{\rm B}$ states. These datasets are further separated for positive (top panel) and negative (middle panel)	$p_{\rm miss}$. They follow a $\sin \phi_{pq}$  dependence with a positive amplitude for the J$^{\pi}$ = 3/2$^{-}$ states and an opposite sign for the $J^{\pi}$ = 1/2$^{-}$ excitation in $^{11}$B.  They show a large enhancement in the low-$p_{\rm miss}$ region (setting A) in comparison to high $p_{\rm miss}$ (setting G). Furthermore, the asymmetries of protons with  $p_{\rm miss} < 0$ which corresponds to parallel kinematics are significantly higher than those with $p_{\rm miss} > 0 $ (anti-parallel kinematics). 

For the high-$p_{\rm miss}$ measurement the resolution in the missing-energy spectrum was insufficient to successfully separate the contributions of the radiative tails from the rest of the events in the excitation regions. Since the differences in the asymmetries in these transitions can be large, only the asymmetry with the transition to $^{11}$B ground state could be reliably determined and is shown at the bottom of Fig.~\ref{fig:phip}.
	
\begin{figure}[b!]
\center{\large $^{12}{\rm C}(\vec e,e'p)$}
	\includegraphics[width=\columnwidth]{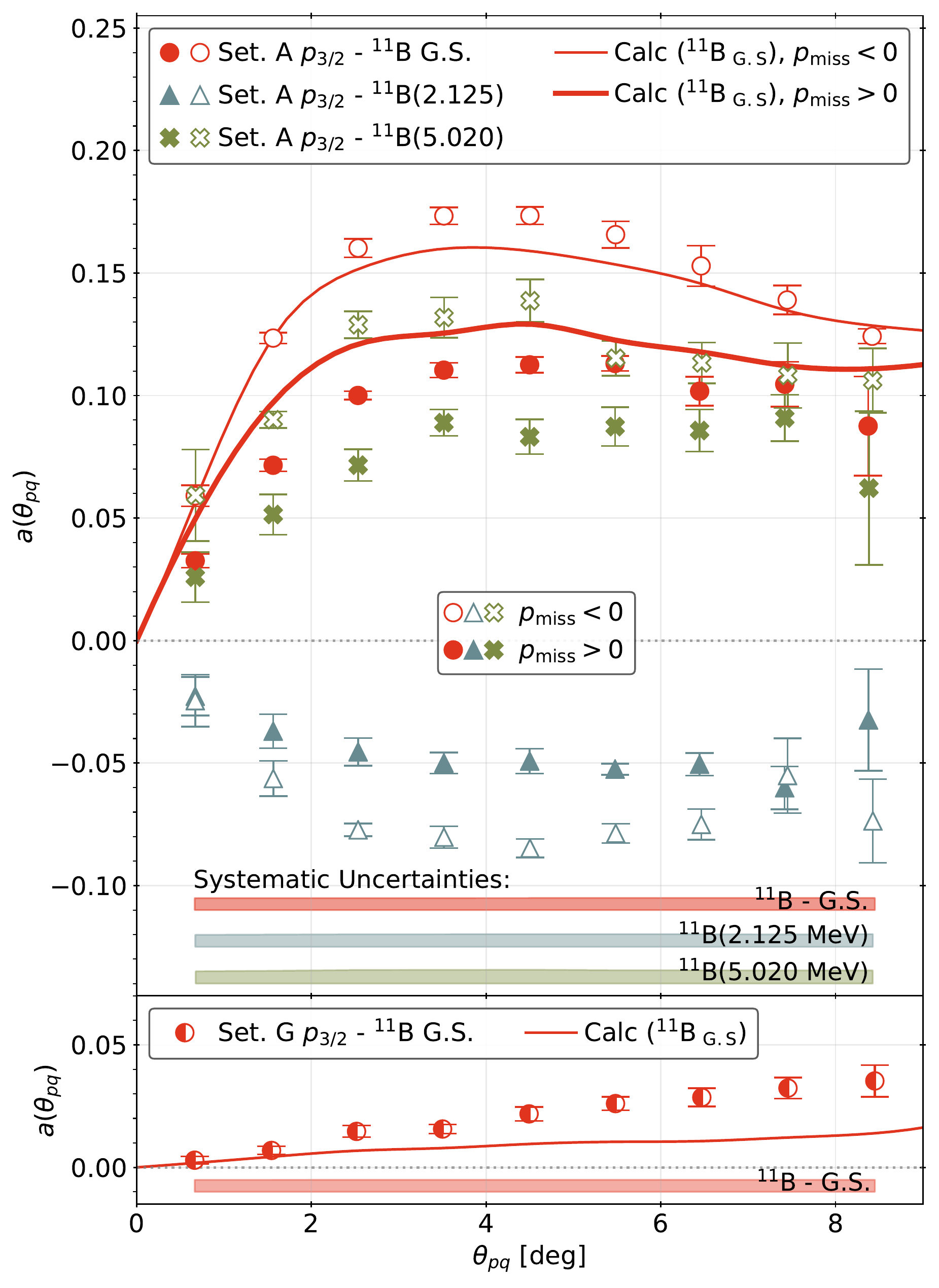}
	\caption{
	    Same as Fig.~\ref{fig:theta} for knock-out from $p$-shell in $^{12}$C. Top panel is showing amplitudes for three different final states at positive and negative $p_{\rm miss}$ obtained at low $p_{\rm miss}$ setting (setting A), while the lower panel shows amplitudes for transition to $^{11}\mathrm{B}$ ground state obtained at high (negative) $p_{\rm miss}$ setting (setting G).
	} 
\label{fig:theta_p} 
\end{figure}

\subsection{$\theta_{pq}$ angular dependence}
In order to examine the dependence of the helicity asymmetry on the scattering angle of the knock-out proton, $\theta_{pq}$, we partitioned the data into slices by $\theta_{pq}$.  The $\phi_{pq}$ dependence of $A$ within each slice was found to be consistent with a sine shape. The amplitude of the sine shape was extracted by fitting the helicity asymmetry within each slice as  
\begin{equation}
A=a(\theta_{pq})\sin(\phi_{pq})
\end{equation}
where $a(\theta_{pq})$ is the fitted amplitude.

These amplitudes, plotted as functions of $\theta_{pq}$, for $^2$H (top panel) and for $s$ protons in $^{12}$C (bottom panel) are shown in Fig.~\ref{fig:theta}. The amplitudes for the $p$ protons in $^{12}$C are shown in Fig~\ref{fig:theta_p}. The amplitudes $a(\theta_{pq})$ tend to vanish when extrapolating $\theta_{pq}$ to zero, as expected, regardless of the nucleus, the kinematic setting, or the shell that the proton is knocked out from.  For the $^2$H data, the amplitude $a(\theta_{pq})$ becomes increasingly negative with increasing $\theta_{pq}$. The measured asymmetry for the $s$-shell protons in $^{12}$C is consistent with zero within the systematic uncertainties and the dependence on the scattering angle cannot be determined. The indication of a negative amplitude in the high-$p_{\rm miss}$ region is hardly significant.  

For the $p$-shell $^{12}$C measurements at low-$p_{\rm miss}$ , the amplitudes are positive for the transitions to the $^{11}$B ground state and the 5.020 MeV excitation (J$^{\pi}$=3/2$^-$), and negative for the 2.125 MeV excitation (J$^{\pi}$=1/2$^-$). The overall $\theta_{pq}$-dependent behavior is considerably different for the two settings.  At Setting A, where $p_{\rm miss}$ is small, the helicity asymmetry is comparatively large (for all transitions), with $a(\theta_{pq})$ peaking at $\theta_{pq}\approx 4\degree$, and decreasing at larger $\theta_{pq}$.  In the large-$|p_{\rm miss}|$ sector (setting G), the amplitude for the transition to the $^{11}$B ground state increases within our measured range of $\theta_{pq}$ (up to $\approx8\degree$), with a nearly constant slope, (reaching $\approx 0.02$).
The difference between the positive and negative $p_{\rm miss}$ sectors is clearly reflected also in the $\theta_{pq}$ dependence.

\subsection{Missing-momentum dependence ($^{12}\mathrm{C}$)}
\begin{figure}[ht!]
	\center{\large $^{12}{\rm C}(\vec e,e'p)^{11}{\rm B}_{\rm G.S.}$;~~Set. A $p_{3/2}$}
	\includegraphics[width=\columnwidth]{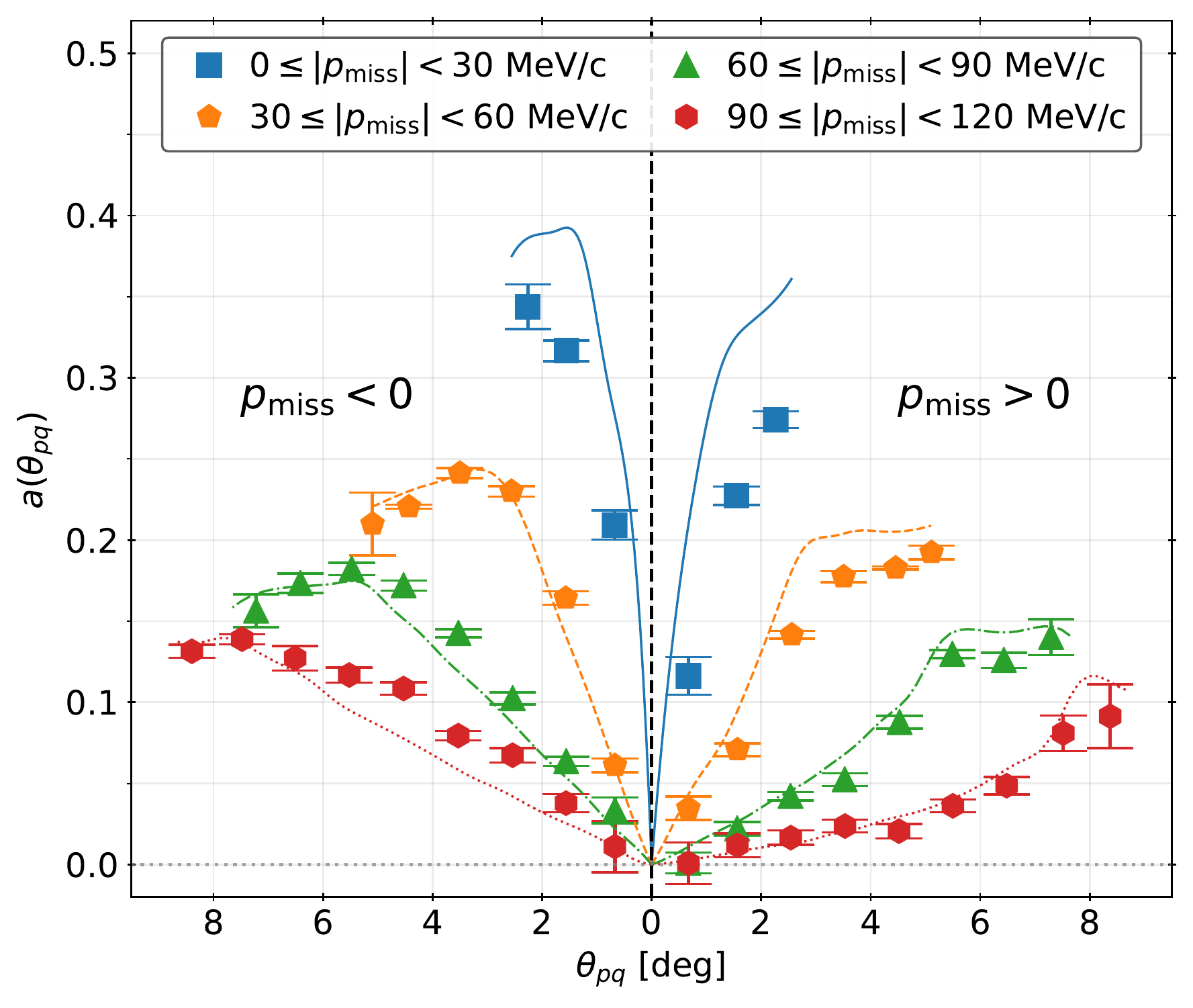}
	\caption{
		Fitted amplitude $a(\theta_{pq},p_{\rm miss})$ of the helicity asymmetry as binned by $\theta_{pq}$ for $p$-shell protons knocked out of $^{12}$C at the low-$|p_{\rm miss}|$ (setting A). To obtain fits only data with the residual nucleus left in the ground state was used. Different symbols (colored online) represent different bins in $|p_{\rm miss}|$. The curves represent the calculations using the full optical potential.
	}
	\label{fig:pmiss_theta}
\end{figure}

The measured asymmetries for the protons extracted from the $p$-shell in $^{12}$C with the transition to  $^{11}$B ground state show a strong variation between the large-$p_{\rm miss}$ data where the measured asymmetry is relatively small, and the low-$p_{\rm miss}$ region where they are much larger. To study the $p_{\rm miss}$ dependence of the asymmetry, we divided the data of the low-$|p_{\rm miss}|$ region (setting A) into bins of $|p_{\rm miss}|$ and repeated the study of the $\theta_{pq}$ angular dependence within each bin, extracting the amplitude, $a(\theta_{pq})$ of the sine-shaped dependence. The angular dependence of the amplitudes in each $p_{\rm miss}$ bin is shown in Fig.~\ref{fig:pmiss_theta}. The asymmetries increase with $p_{\rm miss}$ approaching zero.  

Enhancements of the asymmetries measured for $p$-shell protons around $p_{\rm miss}\approx0$   in $^{12}$C have been observed also in the transferred polarization in the $(\vec e,e' \vec p)$ \cite{ceepLet,ceepComp} as well as in the induced asymmetry in the $(e,e' \vec p)$ reactions \cite{induced}. The polarization-transfer double ratio $(P'_x/P'_z)^{^{12}C_p}/(P'_x/P'_z)^{^1\rm H}$ was  high compared to other nuclei and  shells when compared at the same virtuality \cite{ceepLet}. Also, the induced polarization is considerably larger compared to other measurements \cite{induced}.  Each of these effects is consistent with theoretical calculations (described below in Sec.~\ref{sec:calc}). They have all been attributed to large FSI effects in this region near the minimum of the $p$-shell momentum distribution.  

\section{Comparison to calculations}
\label{sec:calc}
The helicity asymmetries measured for $^2$H and $^{12}$C have been compared with the results of theoretical predictions. For $^2$H we have used a non-relativistic calculation \cite{Arenhovel} including a realistic $NN$-potential, meson-exchange (MEC) and isobar currents (IC), and relativistic contributions (RC) of leading order. For the bound and scattering states, the realistic Argonne $V_{18}$ potential \cite{PhysRevC.51.38} has been taken.  As nucleon electromagnetic form factors, we used the parameterizations from  \cite{Bernauer}. The calculations were obtained for sub-sets of events over the entire kinematic phase space and considering the statistics in each kinematical bin, thus reflecting an average over data in the bin.

The calculations for the $^2$H predict the helicity asymmetry very well, both as a function of $\phi_{pq}$ and $\theta_{pq}$, as shown in the top set of panels in Figs.~\ref{fig:phid} and \ref{fig:theta}, respectively (solid curves).    

It was previously observed that the spin-orbit ($L\cdot S$) part of the final-state interactions is primarily responsible for the non-zero values of the {\it induced} polarization in quasi-free $(e,e'\vec p)$ scattering \cite{induced,Woo}.  In order to determine the role of the $L\cdot S$ interaction in the   helicity asymmetry, we performed the calculations again with the $L\cdot S$ part of the $NN$ potential switched off (dashed curves in Fig.~\ref{fig:phid}).  We found that the no-$L\cdot S$ calculations yield $A\approx 0$, implying that the helicity asymmetry for $^2$H is primarily due to the $L\cdot S$ part of the $NN$ potential.

For $^{12}$C, calculations were performed  using a program \cite{Meucci:2001qc} based on the relativistic distorted-wave approximation (RDWIA) where the FSI between the outgoing proton and the residual nucleus are described by a phenomenological relativistic optical potential. The original program \cite{Meucci:2001qc} was modified \cite{ceepTim} in order to account for non-coplanar kinematics, by including all relevant structure functions \cite{Boffi:1996ikg}.  In the RDWIA calculations, only the one-body electromagnetic nuclear current is included. We chose the current operator corresponding to the cc2 definition \cite{DEFOREST1983232}, and we used the same parametrization of the nucleon form factors \cite{Bernauer} as in the $^2$H calculations. The relativistic  proton bound-state wave functions for the $p_{3/2}$ and $s_{1/2}$ states were obtained from the NL-SH parameterization  \cite{SHARMA1993377} and the scattering states from the so-called  ``democratic'' parameterization of the optical potential \cite{PhysRevC.80.034605}. Even though the optical potential is an important ingredient of the calculation, we did not observe a significant difference in results when we repeated the calculation using the EDAI \cite{Cooper:1993nx} potential for $^{12}\mathrm{C}$. 

We note that the calculations assume a discrete state for the residual nuclear system, a pure $p_{3/2}$ and $s_{1/2}$ hole in the target. For the $s$-shell knockout the residual nuclear system belongs in reality to a continuum spectrum of states. For the $p$-shell knockout the missing-energy spectrum in Fig.~\ref{fig:emiss_histogram} shows that the knockout of $p$-shell protons leads to several final states. Observation of states other than $\mathrm{J}^{\pi}$ =$3/2^{-}$ indicate that there are correlations in the $^{12}$C target nucleus which are not contained in the model wave-functions. This confirms the results of previous experiments and is a clear indication of correlations which are not included in the present calculations. For the description of the three observed final states the model would require the overlap functions from calculations of the hole spectral function. The present RDWIA calculations assume that the residual nucleus is a $3/2^{-}$ state, a $p_{3/2}$ hole state, that can be identified with the $^{11}B$ ground state.

These calculations are shown as solid curves in Figs.~\ref{fig:phis},~\ref{fig:phip},~\ref{fig:theta},
and \ref{fig:theta_p}. In Fig.~\ref{fig:pmiss_theta}, they are shown as curves of varying styles, corresponding to different slices in $p_{\rm miss}$. For the $s$-shell knockout, the calculations predict a much larger negative amplitude than observed in the data. For the $p$-shell (Figs.~\ref{fig:phip} and \ref{fig:theta_p}) there is a general agreement between calculations and the measurements with small deviations. At low-$p_{\rm miss}$ the calculation slightly underestimates the measured asymmetries with $p_{\rm miss} <0$ and overestimates those with $p_{\rm miss} > 0$. Moreover, one may note that the calculation predicts a modulation of the sine dependence in the high-$p_{\rm miss}$ region, which is absent in the measured data. It is interesting to note the comparison of the calculation to the data in the different $p_{\rm miss}$ bins which are shown in Fig.~\ref{fig:pmiss_theta}: overestimating at lower $p_{\rm miss}$ and underestimating at higher $p_{\rm miss}$. This is consistent with the comparison in the high-$p_{\rm miss}$ measurement (Set. G in Fig.~\ref{fig:theta_p}).

Performing the calculations with the $L\cdot S$ interaction switched off (dashed curves) yields a very small helicity asymmetry for $s$-shell knockout, closer to the data than the full-potential calculations.  A possible explanation for this is that the theory overestimates the contribution from the $L\cdot S$ interaction. However, it is also possible that the non-$L\cdot S$ contributions are in reality larger in magnitude than in the present calculation and cancel out the $L\cdot S$ contributions, leaving only a small helicity asymmetry.  

For the $p$-shell knockout, the no-$L\cdot S$ calculations yield a helicity asymmetry that is sizeable compared to the measured values, implying that the $p$-shell knockout is sensitive to both $L\cdot S$ and non-$L\cdot S$ effects.  Furthermore, the $p$-shell calculations for Setting G with the $L\cdot S$ switched off are closer to the data than the full-potential calculations.

\section{Conclusions}
We presented new measurements of the electron-helicity asymmetry on $^2$H and $^{12}$C. The quality
of the new data allows a meaningful comparison with state-of-the-art calculations. The data from
different shells, and different residual nuclear states, and measured over a large range of $p_{\rm miss}$ provide a test of the models of the reaction mechanisms and highlight ingredients in the calculations which may not be properly described. The overall comparison indicates that the reaction mechanisms are described well in $^2$H and the $p$-shell protons in $^{12}$C. However, discrepancies between the calculations for the $s$-shell proton in $^{12}$C and the measured data point to the need for a better treatment of the continuous spectrum of the  states as well as the description of the target ground state in the calculation.

Our data show also the significance of covering the kinematic phase space of both $p_{\rm miss}$ and the
angular dependence of the knocked-out proton (with respect to the momentum transfer, i.e. $\theta_{pq}$), in helicity-asymmetry measurements. They may have a different sensitivity to various ingredients to the
structure function: its dependence on the momentum distribution of the proton which is
determined by its wave-function, and the reaction mechanism.

\section{Acknowledgements}
We would like to thank the Mainz Microtron operators and technical crew for the excellent operation of the accelerator. This work is supported by the Israel Science Foundation (Grants 390/15, 951/19) of the Israel Academy of Arts and Sciences, by the Israel Ministry of Science, Technology and Spaces, by the PAZY Foundation (Grant 294/18), by the Deutsche Forschungsgemeinschaft (Collaborative Research Center 1044), by the U.S. National Science Foundation (PHY-1812382), by the  United States-Israeli Binational Science Foundation (BS) as part of the joint program with the NSF (grant 2017630), and by the Croatian Science Foundation Project No. 1680. We acknowledge the financial support from the Slovenian Research Agency (research core funding No.~P1\textendash 0102).  
\FloatBarrier
\bibliographystyle{elsarticle-num}

\addcontentsline{toc}{section}{\refname}\small{\bibliography{hdep}}
\clearpage

\end{document}